\documentclass[aps,prl,amsmath,amssymb,twocolumn,superscriptaddress,showpacs,floatfix]{revtex4-2}
\usepackage[T1]{fontenc}
\usepackage{graphicx}
\usepackage{dcolumn}
\usepackage{bm}
\usepackage{layout}
\usepackage{lipsum}
\usepackage{amsmath}
\usepackage{physics}
\usepackage{xcolor}
\usepackage{longtable}
\usepackage{placeins}
\usepackage{booktabs}

\usepackage[colorlinks=true, citecolor=blue, linkcolor=blue, urlcolor=blue]{hyperref}

\begin{document}

\preprint{APS/123-QED}

\title{Microscopic Emergence of Ancilla Lattice Physics in the Emery Model}

\newcommand{\affiliationUW}{
Institute of Theoretical Physics, Faculty of Physics, University of Warsaw, Pasteura 5, PL-02093 Warsaw, Poland
}
\newcommand{\affiliationTUM}{Physics Department, TUM School of Natural Sciences, Technical University
of Munich, Garching, Germany.}

\newcommand{\sx}[1]{s_{x,#1}}
\newcommand{\sy}[1]{s_{y,#1}}
\newcommand{\aj}[2]{a_{#1,#2}}
\newcommand{\pind}[1]{{\textbf{r}+\frac{\textbf{#1}}{2},\sigma}}
\newcommand{\ta}{\tilde\alpha}
\newcommand{\tb}{\tilde\beta}
\newcommand{\td}{\tilde d}

\author{Mikołaj Walicki}
\affiliation{\affiliationUW}
\author{Johannes Knolle}
\affiliation{\affiliationTUM}

\author{Krzysztof Wohlfeld}
\affiliation{\affiliationUW}
\email{}
\date{\today}

\begin{abstract}

We derive a controlled low-energy effective theory for the three-band Emery model on the Lieb lattice near (n=2) filling. Starting from the positive charge-transfer regime, a sequence of Schrieffer--Wolff transformations rigorously establishes a direct microscopic mapping onto an ancillary lattice structure. For realistic cuprate-like parameters, this framework naturally reproduces the conventional Fermi-liquid (FL) phase at a filling shifted by one electron per site. Our analysis identifies the precise microscopic conditions required to access the exotic fractionalized Fermi-liquid (FL$^*$) phase. We show that realizing FL$^*$
 requires nonstandard cuprate parameter regimes together with additional interactions that stabilize a quantum spin liquid in the background ancilla layer. These results establish a microscopic foundation for ancilla physics in multiorbital materials and clarify the conditions for realizing FL$^*$. We discuss the possibility to replicate ancilla model physics in Lieb lattice cold atom simulations.

\end{abstract}

\maketitle

{\it Introduction---}The recently proposed ancilla framework provides an appealing route toward realizing metallic states beyond the conventional Landau paradigm~\cite{Zhang_2020}. By coupling itinerant electrons to two additional localized degrees of freedom, the ancilla model naturally hosts both ordinary Fermi liquids (FL) and fractionalized Fermi liquids (FL*) with reduced Fermi volume coexisting with fermionic spinon excitations~\cite{Mascot_2022, Muller_2025}. As such, it offers a minimal setting in which to investigate the interplay between strong correlations, emergent gauge structures, and unconventional metallic behaviour.

A major challenge is to identify microscopic systems in which the ingredients of the ancilla construction arise naturally rather than being introduced as an effective theory. One obvious candidate is the physics of the high-temperature superconducting cuprates~\cite{Bednorz_1986, Bonetti_2026, Christos_2024}. Indeed, the ancilla lattice model supports a transition between FL and FL* phases resembling those observed experimentally in the cuprates~\cite{Keimer_2015}. Nevertheless, the connection remains indirect. First, it relies on a highly nontrivial mapping from the single-band Hubbard model with a doubling of the local Hilbert space~\cite{Bonetti_2026}. Second, although the Hubbard model is widely believed to capture the essential low-energy physics of the cuprates, its quantitative applicability to these materials remains the subject of ongoing debate~\cite{Chen_2013,Ebrahimnejad_2014, Martinelli_2024, Vucicevic_2026}.

\begin{figure}
    \centering
    \includegraphics[width=0.9\linewidth]{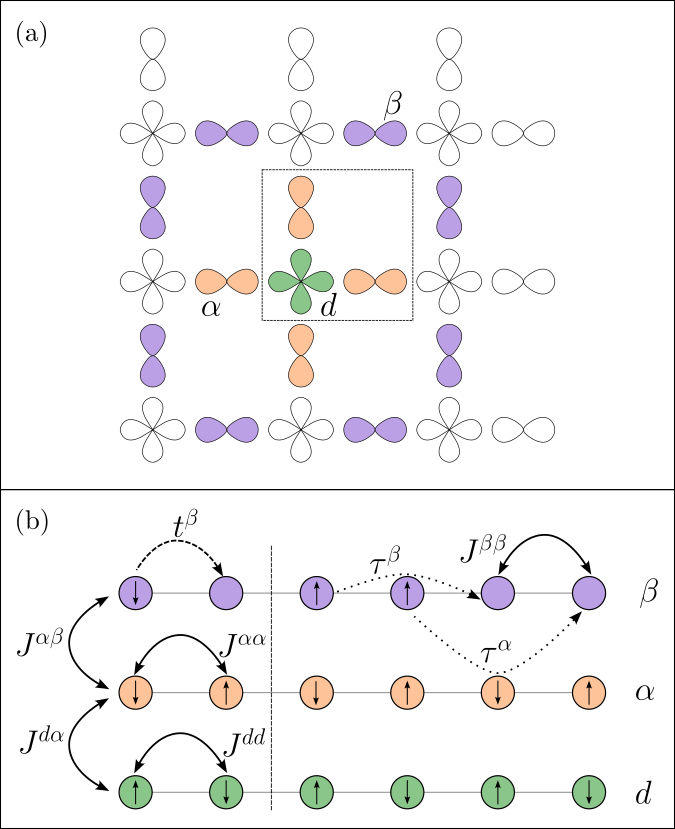}
    \caption{(a) Lieb lattice representation of the CuO$2$ plane, with Cu $d{x^2-y^2}$ orbitals (green) and O $p_x,p_y$ orbitals combined into $\alpha$ (orange) and $\beta$ (violet) states (dominant contributions shown). (b)~Comparison between the ancilla lattice model~\cite{Bonetti_2026} and the effective low-energy model derived here. Both contain three coupled layers: itinerant electrons in the top layer and two spin layers with antiferromagnetic interactions below. The present model solely differs by additional interaction terms, including a $t-J$ description of the itinerant layer and three-site processes.
    Terms common to both models are grouped to the left of the vertical guideline.
    }
    \label{fig:fig1}
\end{figure}

Here, we explore a different route to ancilla model physics based on the  three-band Emery model~\cite{Emery_1987, Weber_2014}---the canonical microscopic model for the high-temperature superconducting cuprates~\cite{Zaanen_1985}, see Fig.~\ref{fig:fig1}(a). The intrinsic multiorbital and strongly correlated character of this system, comprising one Cu and two O orbitals per unit cell of the Lieb lattice, naturally assigns distinct orbital character to the itinerant and ancillary degrees of freedom, as illustrated in Fig.~\ref{fig:fig1}(a).

\FloatBarrier
Inspecting Fig.~\ref{fig:fig1}(b), one quickly realizes that ancilla physics can arise in this setting only at a filling of $n=2+x$ electrons per unit cell. At $n=2$, two of the three available orbitals have their lower Hubbard sectors fully occupied, leaving the third orbital available to accommodate additional carriers. A small electron doping, parametrised by $x$, then generates a metallic sector coupled to a correlated background that naturally realizes the ancilla degrees of freedom. This well-defined limit of the Emery model is distinct from the conventional cuprate phenomenology, but provides a setting in which to investigate the microscopic origin of the ancilla construction.

The natural question is then whether this doped Emery model faithfully reproduces the phenomenology of the ancilla theory and, in particular, whether it supports the two characteristic metallic phases of the ancilla phase diagram, FL and FL$^*$. In this work, we show that the three-band Emery model at filling $n=2+x$ can be rigorously mapped onto the ancilla lattice model. While this filling and the corresponding parameter regime are distinct from those relevant to the cuprates themselves, the resulting effective theory naturally reproduces the conventional FL and identifies the microscopic conditions required to realize the FL$^*$.

{\it Model and basis.---}Our starting point is the original three-band Emery model, the canonical model describing the dynamics of interacting electrons on a Lieb lattice.
As this model was first introduced to describe holes doped into the Mott insulating CuO$_2$ plane of the high-temperature superconducting cuprates~\cite{Emery_1987, Jefferson_1992}, below we follow the notation of these foundational papers. The model Hamiltonian then reads:
\begin{widetext}
\begin{equation}
    \label{eq:H_Emery}
    \begin{aligned}
        \mathcal H_{dp}&=\sum_{\textbf{r},\sigma}\left\{ t_{pd}[d_j^\dagger(-p_{\pind{x}}+p_\pind{y}+p_\pind{-x}-p_\pind{-y})+h.c.]\right.\\
        &\left.+t_{pp}(p_\pind{y}^\dagger p_\pind{x}+p_\pind{-y}^\dagger p_\pind{-x}-p_\pind{y}^\dagger p_\pind{-x}-p_\pind{-y}^\dagger p_\pind{x}+h.c.)\right.\\
        &\left.+\Delta(n^p_{\pind{x}}+n^p_{\pind y})+\frac 12Un_{\textbf{r},\sigma}n_{\textbf{r},\Bar\sigma}+\frac12U_{pp}(n^p_{\textbf{r},x,\sigma}n^p_{\textbf{r},x,\Bar\sigma}+n^p_{\textbf{r},y,\sigma}n^p_{\textbf{r},y,\Bar\sigma})\right\},
    \end{aligned}
\end{equation}
\end{widetext}
with $d_\textbf{r}^\dag$, $p_\textbf{r}^\dag$ creating a hole on the $d$ or $p$ orbital respectively on a site located at position \textbf{r}, and $x,y$ enumerating the two different positions of $p$ orbitals in the Lieb lattice unit cell, {\it cf}.~Fig.~\ref{fig:fig1}(a).

The above Hamiltonian is essentially an extended Hubbard model with three types of orbitals: $d$, $p_x$ and $p_y$. The considered hoppings are either between the nearest neighbor $d$ and $p$ orbitals ($t_{pd}$) or between the $p$ orbitals ($t_{pp}$), with no direct $d-d$ hopping, and the differing signs caused by the relative phases of the orbitals. The $p$ orbitals are higher in energy (in the hole language used here), with the difference $\Delta$ called the charge transfer energy. There is also an on-site Hubbard interaction on the $d$ and $p$ orbitals, with respective strengths $U$ and $U_{pp}$. 

\begin{figure*}[ht!]
    \centering
    \includegraphics[width=\linewidth]{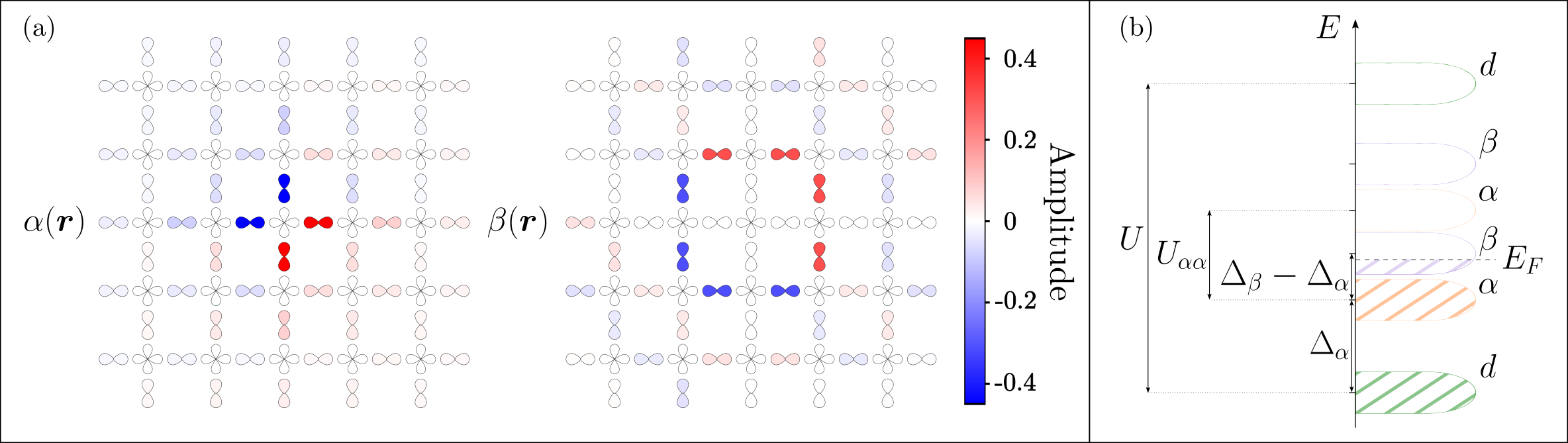}
    \caption{
    (a) Decomposition of $\alpha (\textbf{r} )$ and $\beta (\textbf{r})$ orbital in the $\{d, p_x, p_y \}$ basis, respectively. The colours denote the contribution of a given orbital towards the $\alpha$ or $\beta$ orbital respectively. The $d$ orbitals are left blank as they do not contribute to the $\alpha$ and $\beta$ orbitals. The convention for the $\beta$ orbitals is taken from~\cite{Jefferson_1992}, along with their zero weight at the `origin' site $\textbf{r} = 0$.
    (b) Schematic representation of the hierarchy of energy scales required for the derivation of the effective three-band model, see text for more details.}
    \label{fig:fig2} 
\end{figure*}

The above Hamiltonian can be transformed into a different basis that takes into account the symmetry of lattice. This is beneficial because it allows for separation of the $p$ orbitals into two distinct bands, one of which is decoupled from the $d$ band. Probably the most straightforward choice would be to form molecular orbitals from $p$ orbitals on a plaquette formed by one $d$ orbital and four $p$ orbitals (in the cuprate case this would be a  CuO$_4$ cluster): $\eta_1=p_{x/2}+p_{y/2}-p_{-x/2}-p_{-y/2}$ and $\eta_2=p_{x/2}-p_{y/2}-p_{-x/2}+p_{-y/2}$. These, however, are not mutually orthogonal due to the fact that all $p$ orbitals are shared between two plaquettes. One solution is going to momentum space and forming so-called canonical fermions $\alpha$ and $\beta$~\cite{Shastry_1989}:
\begin{equation}
    \label{eq:alpha_beta_definition}
    \begin{aligned}
        \alpha_{\textbf{k},\sigma}&=i(\sx{\textbf{k}} p_{\textbf{k},x,\sigma}-\sy{\textbf{k}}p_{\textbf{k},y,\sigma})/\mu_k,\\
        \beta_{\textbf{k},\sigma}&=-i(\sy{\textbf{k}} p_{\textbf{k},x,\sigma}+\sx{\textbf{k}} p_{\textbf{k},y,\sigma})/\mu_k,
    \end{aligned}
\end{equation}
where $\sx{\textbf{k}}=\sin(\textbf{kx}/2)$, $\sy{\textbf{k}}=\sin(\textbf{ky}/2)$ and $\mu_k=\sqrt{\sx{\textbf{k}}^2+\sy{\textbf{k}}^2}$, with $\textbf{x}=(1,0)$, $\textbf{y}=(0,1)$ being the lattice vectors. These orbitals are also used for the Zhang-Rice singlet construction~\cite{Zhang_1988}. The $\alpha$ orbitals transform as $B_1$ and $\beta$ as $A_1$ under the lattice point group~\cite{Feiner_1996}. Unlike $\eta_{1,2}$ the $\alpha$ and $\beta$ orbitals are orthonormal, but at a cost of no longer being local. Fortunately, as shown in Fig.~\ref{fig:fig2}(a), the non-local `tails' vanish quite rapidly.
In fact, most of the weight is concentrated near the centre of the orbital for the $\alpha$ orbital, while the $\beta$ orbital has zero on-site weight and is mostly present on the neighbouring sites, {\it cf}. Fig.~\ref{fig:fig2}(a).

We rewrite the original Hamiltonian in the new $\{ d, \alpha, \beta \}$, basis:
\begin{widetext}
\begin{equation}
\label{eq:full_Hamiltonian}
\begin{aligned}
    \mathcal{H}_{d\alpha\beta}&=\sum_{\textbf{r},\sigma} \Delta(\alpha^\dag_{\textbf{r},\sigma}\alpha_{\textbf{r},\sigma} + \beta^\dag_{\textbf{r},\sigma}\beta_{\textbf{r},\sigma})- 2t_{pd}\sum_{\textbf{r},\textbf{r}',\sigma}(\mu(\textbf{r}-\textbf{r}')d^\dag_{\textbf{r},\sigma}\alpha_{\textbf{r}',\sigma}+H.c.) \\
    &-2t_{pp}\sum_{\textbf{r},\textbf{r}',\sigma}[\nu(\textbf{r}-\textbf{r}')(\alpha^\dag_{\textbf{r},\sigma}\alpha_{\textbf{r}',\sigma} - \beta^\dag_{\textbf{r},\sigma}\beta_{\textbf{r}',\sigma})+\chi(\textbf{r}-\textbf{r}')(\alpha^\dag_{\textbf{r},\sigma}\beta_{\textbf{r}',\sigma}+H.c.)]\\
    &+\sum_{\textbf{r}}\big[U n_{\textbf{r},\uparrow}^dn_{\textbf{r},\downarrow}^d+U_{\alpha\alpha}(n^\alpha_{\textbf{r},\uparrow}n^\alpha_{\textbf{r},\downarrow}+n^\beta_{\textbf{r},\uparrow}n^\beta_{\textbf{r},\downarrow})\big],
\end{aligned}
\end{equation}
\end{widetext}
where the hopping coefficients are given as $\mu(\textbf{r})=N^{-1}\sum_\textbf{k}e^{-i\textbf{kr}}\mu_\textbf{k}$, $\nu(\textbf{r})=N^{-1}\sum_\textbf{k}e^{-i\textbf{kr}}\left(\frac{2s_{x\textbf{k}}s_{y,\textbf{k}}}{\mu_k}\right)^2$, $\chi(\textbf{r})=N^{-1}\sum_\textbf{k}e^{-i\textbf{kr}}\frac{2(s_{x,\textbf{k}}^2-s_{y,\textbf{k}}^2)s_{x,\textbf{k}}\sy{\textbf{k}}}{\mu_k^2}$. These decay quickly with distance $|{\bf r}|$, {\it cf.}~\cite{Jefferson_1992}. We retain only the local part of the Hubbard interaction between the $\alpha$ and $\beta$ orbitals, $U_{\alpha\alpha}$, with detailed derivation in the new basis presented in the SM~\cite{SM}.

{\it Schrieffer-Wolff transformation and effective model.---}Our main goal now is to {\it rigorously} derive a low-energy version of the Hamiltonian Eq.~\eqref{eq:full_Hamiltonian}. To this end we use the Schrieffer-Wolff transformation~\cite{Schrieffer_1966} and assume that the following conditions 
are satisfied (see Ref.~\cite{Bravyi_2011} and~\cite{SM} for the detailed derivation):
\begin{enumerate}
    \item The hole filling has to be sufficiently high to half-fill both the $d$ and the $\alpha$ orbitals and occupy some of the $\beta$ orbitals in order to have itinerant holes. This requires the hole filling per unit cell $n>2$.
    \item All of the hopping terms should be `sufficiently' small for the Schrieffer-Wolff transformation to be well-defined (see~\cite{SM}).
    \item The states must be well separated in energy, as illustrated in Fig.~\ref{fig:fig2}(b). While this hierarchy is motivated by the conventional parameter values for cuprates~\cite{McMahan_1990}, it imposes constraints on the Hamiltonian parameters, requiring
$
U_{\alpha\alpha} =\psi_{0000}U_{pp}
\approx0.21U_{pp}>4\nu(0)t_{pp}\approx2.9t_{pp}
$. Here $\psi_{0000}$ is the coefficient of the dominant term in the expansion of the Hubbard interaction on oxygen atoms, discussed in more details in~\cite{SM}.
\end{enumerate}

\begin{table}
\centering
\setlength{\tabcolsep}{3.5pt}

\begin{tabular}{c c c}
\toprule
\multicolumn{3}{c}{Emery model~\eqref{eq:H_Emery}} \\
\midrule
 & in [$\Delta$] & in [eV] \\
\midrule
$t_{pd}$   & 0.37 & 1.3 \\
$t_{pp}$   & 0.09 & 0.3 \\
$\Delta$   & 1.00 & 3.51 \\
$U$        & 2.59 & 9 \\
$U_{pp}$   & 1.42 & 5 \\
\bottomrule
\end{tabular}
\hspace{0.25cm}
\begin{tabular}{c c c}
\toprule
\multicolumn{3}{c}{Effective model~\eqref{eq:heff}} \\
\midrule
 & in [$J^{d\alpha}$] & in [eV] \\
\midrule
$t^\beta$              & -0.032 & -0.164 \\
$J^{\alpha\alpha}$     & 0.010  & 0.051 \\
$J^{\alpha\beta}$      & 0.005  & 0.025 \\
$J^{d\alpha}$          & 1.000  & 5.104 \\
$J^{dd}$               & 0.002  & 0.008 \\
$\tau^\alpha$          & -0.001 & -0.007 \\
$\bar{\tau}^\alpha$    & -0.001 & -0.005 \\
$\tau^\beta$           & -0.013 & -0.068 \\
\bottomrule
\end{tabular}

\caption{
(Left) Parameters of the original Emery Hamiltonian used in this work. These correspond to a cuprate parameter set~\cite{McMahan_1990} with slight modifications to satisfy the hierarchy of energy scales shown in Fig.~\ref{fig:fig2}(b). Values are given in units of $\Delta$ and in eV. 
(Right) Parameters of the effective model. Values are given in units of $J^{d\alpha}$ and in eV. The listed values correspond to the shortest distance at which the respective couplings do not vanish: on-site for $J^{d\alpha}$, nearest-neighbor for intralayer coefficients, and a `knight's move' $(2,1)$ for the couplings of $\alpha$ and $\beta$ orbitals. All coefficients are small compared with~$J^{d\alpha}$.
}
\label{tab:parameters}
\end{table}
\FloatBarrier
In order to derive an effective multi-band Hamiltonian a series of Schrieffer-Wolff transformations is performed, projecting out the higher energy states. 
Following the energy hierarchy outlined above, we successively perform expansions in the on-site interaction $U$ on the $d$ orbitals, the on-site interaction $U_{\alpha\alpha}$ on the $\alpha$ and $\beta$ orbitals, and finally the charge-transfer energy $\Delta$.
The final low-energy effective Hamiltonian obtained this way reads: 
\begin{subequations}
    \label{eq:H_3band_effective}
    \begin{align}
       H_{\text{eff}}&=\sum_{j,\sigma}\big(\Delta_\alpha \tilde n^\alpha_{j\sigma}+\Delta_\beta\tilde n^\beta_{j\sigma}\big)
       -\sum_{j_1,j_2,\sigma}t^\beta_{j_1j_2}\tilde\beta_{j_1\sigma}^\dag\tilde\beta_{j_2\sigma}\label{subeq:base_H_3band''}\\
            &+\sum_{j_1\neq j_2}J^{\alpha\alpha}_{j_1j_2}\Big(\boldsymbol{s}^\alpha_{j_1}\boldsymbol{s}^\alpha_{j_2}-\frac{1}{4}\Big)+J^{\beta\beta}_{j_1j_2}\Big(\boldsymbol{s}^\beta_{j_1}\boldsymbol{s}^\beta_{j_2}-\frac{\tilde n_{j_1}^\beta\tilde n_{j_2}^\beta}{4}\Big)\label{subeq:J_H''_3band}\\
            &+\sum_{j_1,j_2}\Big[J^{d\alpha}_{j_1j_2}\Big(\boldsymbol{S}_{j_1}\boldsymbol{s}^\alpha_{j_2}-\frac{1}{4}\Big)+J^{\alpha\beta}_{j_1j_2}\Big(\boldsymbol{s}^\alpha_{j_1}\boldsymbol{s}^\beta_{j_2}-\frac{\tilde n_{j_2}^\beta}{4}\Big)\Big]\notag\\
            &+\sideset{}{'}\sum_{j_1,j_2,j_3,\sigma}\bigg[\Big(\tau^\alpha_{j_1j_2j_3}+\bar\tau^\alpha_{j_1j_2j_3}\Big)\tb^\dag_{j_1\sigma}s^{\alpha\Bar\sigma}_{j_2}\tb_{j_3\Bar\sigma}\notag\\
            &-\tau^\alpha_{j_1j_2j_3}\tb_{j_1\sigma}\tilde n_{j_2\sigma}^\alpha\tb_{j_3\sigma}^\dagger-\bar\tau^\alpha_{j_1j_2j_3}\tb^\dag_{j_1\sigma}\tilde n^\alpha_{j_2\Bar\sigma}\tb_{j_3\sigma}\label{subeq:3site_H''_3band}\\
            &+\tau^\beta_{j_1j_2j_3}(\tb^\dag_{j_1\sigma}s^{\beta\Bar\sigma}_{j_2}\tb_{j_3\Bar\sigma}-\tb^\dag_{j_1\sigma}\tilde n^\beta_{j_2\Bar\sigma}\tb_{j_3\sigma})\bigg]\notag\\
&-\sum_{j_1,j_2,\sigma}\tau_{j_1j_2j_1}\tilde n_{j_1\sigma}^\alpha(1-n_{j_2\sigma}^\beta)(1-n^\beta_{j_2\bar\sigma})\label{subeq:interaction_H''_3band},
    \end{align}
\end{subequations}
where the primed sum denotes summation over all combinations of $j_1,j_2,j_3$ in which orbitals of the same type ($\alpha, \beta, d$) are not at the same site. The $\tilde\beta_{j\sigma} = \beta_{j\sigma}(1-n_{j\bar\sigma})$ are the usual constrained annihilation operators, and analogous constraint applies to the $\tilde n_{j\sigma}=n_{j\sigma}(1-n_{j\bar\sigma})$ operators. The parameters $\Delta_\alpha = \Delta - 2t_{pp}\nu(0)$ and $\Delta_\beta=\Delta+2t_{pp}\nu(0)$ are the on-site energies of the $\alpha_{\bf r}$ and $\beta_{\bf r}$ orbitals respectively. The $J$'s are the exchange interactions between sites in layers specified in the superscript, and the $\tau$'s denote the various three-site processes connecting two $\beta$ sites and an intermediary site denoted in the superscript (the bar denotes that the intermediary site has to have an opposite spin). The relations between the parameters of this effective model and the parameters of the original Hamiltonian are given in Ref.~\cite{SM}. 

The Schrieffer--Wolff transformation is performed up to second order, and therefore it does not generate an exchange interaction between the $d$ orbitals, since the latter arises only at fourth order in perturbation theory. This term, which is typically not negligible compared to other fourth-order contributions~\cite{Ebrahimnejad_2014}, can nevertheless be straightforwardly derived and added in the form
\begin{equation} \label{eq:4thorder}
\begin{aligned}
    H_{dd}
    &=
    \sum_{j_1,j_2,j_3}
    J^{dd}_{j_1j_2j_3}
    \,{\bf S}_{j_1}\!\cdot\!{\bf S}_{j_2}
    \\
    &\approx
    J^{dd}
    \sum_{\langle j_1,j_2\rangle}
    {\bf S}_{j_1}\!\cdot\!{\bf S}_{j_2},
\end{aligned}
\end{equation}
where $\langle j_1,j_2\rangle$ denotes summation over nearest-neighbour pairs. The microscopic origin of this exchange interaction, as well as the relation between $J^{dd}$ and the parameters of the mother Hamiltonian, is discussed in detail in~\cite{SM}. Note that, unlike Lau {\it et al.}~\cite{Lau_2011}, we obtain a finite exchange interaction between the $d$ sites because we employ a different perturbative expansion scheme.

Altogether, we obtain an effective ancilla model describing the dynamics of holes in the strongly doped ($n \geq 2$) system:
\begin{equation}\label{eq:heff}
    H_{\rm anc}= H_{\rm eff}+ H_{dd}. 
\end{equation}
with the following individual terms entering the model, {\it cf}. Fig.~\ref{fig:fig1}(b):
\begin{enumerate}
    \item Eq.~(\ref{subeq:base_H_3band''}) describes the energy of holes in the oxygen orbitals (which is a constant value, because the bands do not exchange electrons) and hopping between the $\beta$ orbitals. The hoppings within the $\alpha$ band and between $d$ and $\alpha$ vanish because the low energy sector with respect to $\Delta_\beta$ has as few holes in the $\beta$ orbitals as possible --- this means that each $d$ and $\alpha$ orbital is already occupied by a single hole, rendering hopping without creating double occupancies impossible.
    \item Eq.~(\ref{subeq:J_H''_3band}) is the intra- and inter-band exchange interactions.
    \item Eq.~(\ref{subeq:3site_H''_3band}) describes the three-site processes. The only changes introduced by the expansion in $\Delta_\beta$ are an extra contribution to the $\beta-\alpha-\beta$ process with spin flip on $\alpha$, and an additional transition without spin flip with the same spin on all three orbitals.
    \item Eq.~(\ref{subeq:interaction_H''_3band}) describes a process in which a hole on the $\alpha$ orbital can lower its energy by virtually hopping into a nearby empty $\beta$ orbital.
    \item Finally, Eq. \eqref{eq:4thorder} describes the spin exchange interactions in the $d$ layer.
\end{enumerate}

{\it Comparison with the ancilla model.---}Already a quick look at Fig.~\ref{fig:fig1}(b) shows that the structure of the resulting effective Hamiltonian is remarkably similar to the ancilla layer Hamiltonian introduced by Sachdev {\it et al.} as an effective description of the Hubbard model~\cite{Bonetti_2026}. Let us now examine this correspondence in more detail.

The ancilla model consists of three layers: one `real' layer containing itinerant electrons and two `hidden' (ancillary) layers carrying spin-$1/2$ moments. The top layer is coupled to the middle layer through a Kondo exchange, while the two hidden layers interact via an antiferromagnetic exchange interaction. The interactions within the hidden layers are left largely unspecified, but are assumed to stabilize a quantum spin liquid in the absence of the top layer [{\it cf.} Fig.~\ref{fig:fig1}(b)].

By comparison, our construction yields a $t$--$J$ model in the top $\beta$ layer, coupled to the middle $\alpha$ layer through both antiferromagnetic exchange and three-site terms. Note that unlike the original ancilla model, the `hidden' layers emerge microscopically and correspond to `physical' orbitals, namely the $\alpha$ and $d$ orbitals.

For conventional cuprate parameters~\cite{McMahan_1990}, the effective couplings obtained from the microscopic derivation (Table~\ref{tab:parameters}) place the system firmly in the conventional Fermi-liquid (FL) regime of the ancilla phase diagram~\cite{Bonetti_2026}. In particular, the top $\beta$ layer is only weakly doped, and its intralayer exchange $J^{\beta\beta}$ is found to be negligibly small, suppressing magnetic correlations within this layer and making it analogous to the free-electron layer of the original ancilla model. Meanwhile, the two lower layers form an almost decoupled product state of interlayer singlets, consistent with the FL phase of the ancilla model~\cite{Bonetti_2026}.

Realizing the fractionalized Fermi-liquid (FL$^*$) phase is considerably more demanding. It requires both tuning the microscopic parameters of the Emery model away from the conventional cuprate regime and stabilizing a quantum spin liquid in the ancillary sector. For example, the FL$^*$ regime requires $J^{\alpha\alpha}$ and $J^{\alpha\beta}$ to become much larger than $J^{d\alpha}$, which can be achieved only in a parameter regime distinct from that of conventional cuprates, such as $t_{pp}\gg t_{pd}$~\cite{Bonetti_2026}. Crucially, the bottom layer must develop a quantum spin liquid rather than the N\'eel order that is readily stabilized here. One possible route is to generate sufficiently strong frustrating interactions, such as next-nearest-neighbour antiferromagnetic exchange $J_2$ or ring exchange, within the bottom layer. Such interactions can in principle arise from higher-order virtual processes in the strong-coupling expansion of the Emery model, but incorporating them systematically is beyond the scope of the present work.

{\it Conclusions.---}Hubbard-like models on the Lieb lattice host a remarkably rich variety of correlated quantum phases. The most extensively studied realization is the three-band Emery model near $n=1$ filling~\cite{Emery_1987}, where doping gives rise to unconventional metallic behaviour, including the pseudogap regime and high-$T_C$ superconductivity~\cite{Bednorz_1986, Alloul_1988, Ding_1996, Keimer_2015}. At the opposite end, the $n=3$ filling is characterized by Lieb's theorem~\cite{Lieb_1989}, which guarantees ferrimagnetism in the half-filled bipartite Hubbard model, while doping leads to various incommensurate spin and charge ordered phases~\cite{Gouveia_2015,Tsai_2015, Everts_, Kumar_2017, Lebrat_2026}. These two limits motivate the exploration of the intermediate $n=2$ regime. While recent studies have focused on this filling in the negative charge-transfer regime, where ligand-centred moments can stabilize altermagnetism~\cite{Kaushal_2025, Chang_2026, Li_2025, Ursin_2026}, here we consider the positive charge-transfer regime relevant to the cuprates.

We have shown that the low-energy physics of the Emery model on the Lieb lattice slightly above $n=2$ can be mapped onto an ancilla lattice model. This establishes a microscopic route by which the ancilla paradigm, originally introduced as an effective framework for correlated metallic states with emergent gauge structures~\cite{Zhang_2020}, emerges from a multiorbital Hubbard Hamiltonian. For realistic cuprate-like parameters, the resulting effective theory naturally realizes the conventional FL phase, while we also identify the microscopic conditions required for the emergence of FL$^*$. Our results thus place the ancilla framework on a microscopic footing and delineate its regime of applicability.

Several directions follow from this work. Experimentally, it will be interesting to investigate the ancilla physics identified here using the recently developed Lieb lattice quantum simulators with ultracold atoms~\cite{Lebrat_2026}, or search for suitably doped materials realizing this regime. 
More broadly, the present framework provides a controlled starting point for exploring how multiorbital electronic structure can give rise to both conventional and fractionalized metallic states.

\section{Acknowledgements}
M.W. and K.W. acknowledge the support
of National Science Centre in Poland under Project No. 2024/55/B/ST3/03144. J.K. acknowledges support from the Deutsche Forschungsgemeinschaft (DFG, German Research Foundation) under grants TRR 360 - 492547816, KN1254/1-2, KN1254/2-1 and under Germany’s Excellence Strategy EXC-2111-390814868, as well as the Munich Quantum Valley, which is supported by the Bavarian state government with funds from the High-tech Agenda Bayern Plus. J.K. further acknowledges support from the Imperial-TUM flagship partnership and the Keck foundation.
Collaboration between M.W., J.K.,
and K.W. was supported by the Tandems for Excellence: Visiting Researchers Programme of the University of Warsaw (IDUB UW).

\bibliography{Cuprates_effective_model.bib}
\clearpage
\onecolumngrid

\setcounter{section}{0}
\setcounter{equation}{0}
\setcounter{figure}{0}
\setcounter{table}{0}
\setcounter{page}{1}

\renewcommand{\thesection}{S\arabic{section}}
\renewcommand{\theequation}{S\arabic{equation}}
\renewcommand{\thefigure}{S\arabic{figure}}
\renewcommand{\thetable}{S\arabic{table}}
\renewcommand{\thepage}{S\arabic{page}}

\begin{center}
    \textbf{{\Large Supplemental Material for `Realizing the Ancilla Layer Model on the Lieb Lattice'}\\
    \vspace{0.5cm}{\large by M. Walicki, J. Knolle and K. Wohlfeld}}\\
\end{center}

\vspace{1cm}
\section{Coulomb repulsion between the $p$ orbitals in the $\{d, \alpha, \beta \}$ basis}
\FloatBarrier
The transition from the $\{d,  p_x$, $p_y \}$ to the $\{d, \alpha$, $\beta \}$ basis causes the Hubbard repulsion on the $p$ orbitals to become rather complex, with several classes of terms. In fact, the Hubbard Hamiltonian then reads as follows:
    \begin{equation}
\label{eq:U_alphabeta}
    \begin{aligned}
        \mathcal H_U^p&=U_{pp}\sum_{{\textbf{r}}}(p_{{\textbf{r}},x,\uparrow}^\dag p_{{\textbf{r}},x,\uparrow}p_{{\textbf{r}},x,\downarrow}^\dag p_{{\textbf{r}},x,\downarrow}+p_{{\textbf{r}},y,\uparrow}^\dag p_{{\textbf{r}},y,\uparrow}p_{{\textbf{r}},y,\downarrow}^\dag p_{{\textbf{r}},y,\downarrow})\\
        &=U_{pp}\sum_{\textbf{r}_1, \textbf{r}_2, \textbf{r}_3, \textbf{r}_4}[\psi_{\textbf{r}_1 \textbf{r}_2\textbf{r}_3\textbf{r}_4}(\alpha_{\textbf{r}_1,\uparrow}^\dag\alpha_{\textbf{r}_2,\uparrow}\alpha_{\textbf{r}_3,\downarrow}^\dag\alpha_{\textbf{r}_4,\downarrow}+\beta_{\textbf{r}_1,\uparrow}^\dag\beta_{\textbf{r}_2,\uparrow}\beta_{\textbf{r}_3,\downarrow}^\dag\beta_{\textbf{r}_4,\downarrow})\\
        &+\zeta_{\textbf{r}_1\textbf{r}_2\Bar{\textbf{r}}_3\Bar{\textbf{r}}_4}\alpha_{\textbf{r}_1,\uparrow}^\dag\alpha_{\textbf{r}_2,\uparrow}\beta_{\textbf{r}_3,\downarrow}^\dag\beta_{\textbf{r}_4,\downarrow}+\ldots+\zeta_{\bar{\textbf{r}}_1\bar{\textbf{r}}_2\textbf{r}_3\textbf{r}_4}\beta_{\textbf{r}_1,\uparrow}^\dag\beta_{\textbf{r}_2,\uparrow}\alpha_{\textbf{r}_3,\downarrow}^\dag\alpha_{\textbf{r}_4,\downarrow}],
    \end{aligned}
\end{equation}
where the coefficients $\psi, \zeta$ are given as triple Fourier transforms of:
\begin{equation} \label{eq:Hpp}
    \begin{aligned}
        \psi_{\textbf{k}_1\textbf{k}_2\textbf{q}}&=\frac{\sx{\textbf{k}_1}\sx{\textbf{k}_2}\sx{\textbf{k}_1+\textbf{q}}\sx{\textbf{k}_2-\textbf{q}}+\sy{\textbf{k}_1}\sy{\textbf{k}_2}\sy{\textbf{k}_1+\textbf{q}}\sy{\textbf{k}_2-\textbf{q}}}{\mu_{\textbf{k}_1}\mu_{\textbf{k}_2}\mu_{\textbf{k}_1+\textbf{q}}\mu_{\textbf{k}_2-\textbf{q}}},\\
        \zeta_{\textbf{k}_1\textbf{k}_2\textbf{q}}&=\frac{\sx{\textbf{k}_1}\sy{\textbf{k}_2}\sx{\textbf{k}_1+\textbf{q}}\sy{\textbf{k}_2-\textbf{q}}+\sy{\textbf{k}_1}\sx{\textbf{k}_2}\sy{\textbf{k}_1+\textbf{q}}\sx{\textbf{k}_2-\textbf{q}}}{\mu_{\textbf{k}_1}\mu_{\textbf{k}_2}\mu_{\textbf{k}_1+\textbf{q}}\mu_{\textbf{k}_2-\textbf{q}}},
    \end{aligned}
\end{equation}
and bars in $\zeta_{\textbf{r}_1\textbf{r}_2\Bar{\textbf{r}}_3\Bar{\textbf{r}}_4}$ mark which indices refer to $\beta$ orbitals. The $\ldots$ in the $\zeta$ term im Eq.~\eqref{eq:Hpp} is a shorthand to represent all orderings of two $\alpha$ and two $\beta$ operators. The terms with odd number of $\beta$ orbitals disappear due to symmetry constraints. The coefficients are symmetric under the permutation of the indices (including bars), as well flipping all the bars. 
\begin{table}
    \centering
    \begin{tabular}{|c|c|c|c|c|}
    \hline
    $\textbf{r}_1$, $\textbf{r}_2$, $\textbf{r}_3$, $\textbf{r}_4$&$\psi_{\textbf{r}_1 \textbf{r}_2\textbf{r}_3\textbf{r}_4}$&$\zeta_{\textbf{r}_1\textbf{r}_2\Bar{\textbf{r}}_3\Bar{\textbf{r}}_4}$&$\zeta_{\textbf{r}_1\bar{\textbf{r}}_2\Bar{\textbf{r}}_3{\textbf{r}}_4}$&$\zeta_{\textbf{r}_1\bar{\textbf{r}}_2\textbf{r}_3\Bar{\textbf{r}}_4}$\\
    \hline
{\bf 0}, {\bf 0}, {\bf 0}, {\bf 0}&0.2109&0.1710&0.0018&0.0027\\
{\bf 0}, {\bf 0}, {\bf 0}, \textbf{x}&-0.0303&-0.0242& 0.0036&0.0053\\
{\bf 0}, {\bf 0}, \textbf{x}, \textbf{x}&0.0590&0.0624&-0.0004&-0.0008\\
\hline
    \end{tabular}
\caption{Values of the parameters $\psi$ and $\zeta$ for different sets of indices $\textbf{r}_1 \textbf{r}_2\textbf{r}_3\textbf{r}_4$. {\bf 0} denotes the zero vector and {\bf x} denotes the lattice vector in the x direction (1, 0). The $\psi_{0000}$ coefficient (responsible for the on-site intraorbital Hubbard-like terms) is the largest, followed by the $\zeta_{00\bar0\bar0}$ coeffecient which gives a $(n^\alpha_{\textbf{r}\uparrow}n^\beta_{\textbf{r}\downarrow}+n^\beta_{\textbf{r}\uparrow}n^\alpha_{\textbf{r}\downarrow})$ term, which can also be included in the effective Hamiltonian without affecting the perturbative expansion. The remaining terms are at least one order of magnitude smaller.}
    \label{tab:psi}
\end{table}
The $\psi$ terms represent the Coulomb repulsion on orbitals of the same type and $\zeta$ terms represent the processes coupling $\alpha$ and $\beta$ orbitals, namely Coulomb repulsion, spin exchange and correlated pair hopping. 

Crucially, as shown in Table~\ref{tab:psi} the on-site ($\textbf{r}_1=\textbf{r}_2=\textbf{r}_3=\textbf{r}_4$) coefficients are the largest in magnitude, with the intraorbital Hubbard-like contribution $\psi_{0000}U_{pp}(n_{\textbf r\uparrow}^\alpha n_{\textbf r\downarrow}^\alpha+n_{\textbf r\uparrow}^\beta n_{\textbf r\downarrow}^\beta)$ dominating. That is why this is the only term listed in the main text and considered when performing the Schrieffer-Wolff transformations. Nevertheless, the $\zeta_{00\bar0\bar0}$ coefficient, which gives a $(n^\alpha_{\textbf{r}\uparrow}n^\beta_{\textbf{r}\downarrow}+n^\beta_{\textbf{r}\uparrow}n^\alpha_{\textbf{r}\downarrow})$ term is also significant, and can be added to the effective Hamiltonian without affecting the perturbative expansions.

\section{Derivation of the Schrieffer-Wolff transformation}

\label{app:SW-transformation}
The form of the Schrieffer-Wolff transformation used in this letter is based on~\cite{Howczak_,Bravyi_2011}.
To begin we need to separate the Hamiltonian into two parts:
\begin{equation}
    \mathcal H=\mathcal H_{diag} + V,
\end{equation}
where the spectrum of $\mathcal H_{diag}$ is exactly known, and its spectrum is divided into groups of states (called energy sectors) separated by an energy gap $\Delta$, and $V$ is a small perturbation. In this letter $\mathcal H_{diag}$ is the part of the Hamiltonian diagonal in number operators, and $V$ constitutes the various hoppings. In order for this division to hold in the full Hamiltonian $\mathcal H$ the perturbation $||V||<\frac\Delta2$. 
The projection operator onto the ground sector of $\mathcal H_{diag}$, $P_0$, is defined in a usual way -- just as the projectors $P_s$ onto sectors with energy $s\Delta$ above the ground sector. These projectors span the whole Hilbert space. The Hamiltonian is then split into two different parts:
\begin{equation}
    \mathcal H = \sum_{r,s}P_r\mathcal HP_s = \sum_{s}P_s\mathcal HP_s +  \sum_{r\neq s}P_r\mathcal HP_s = \mathcal H_0 + \mathcal H_1,
\end{equation}
where $\mathcal H_0$ is the part of the Hamiltonian that does not transfer the state between different energy sectors (block-diagonal), and $\mathcal H_1$ is the part transferring between different energy sectors (block off-diagonal, e.g. $\mathcal H_1$ creates and destroys double occupancies).
Now the goal of the Schrieffer-Wolff transformation is a canonical transformation of the Hamiltonian such that the resulting effective Hamiltonian is block-diagonal. This is most conveniently defined as:
\begin{equation}
    \tilde{ \mathcal H}=e^{-iS}\mathcal He^{iS},
\end{equation}
where $S$ is an operator to be established. This definition can be expanded into a series using the Baker-Campbell-Hausdorff formula:
\begin{equation}
    \tilde{\mathcal H}=\mathcal H - i[S,\mathcal H]-\frac12[S,\mathcal H]_2+\frac i6[S,\mathcal H]_3+\frac{1}{24}[S,\mathcal H]_4+\ldots, 
\end{equation}
    where $[S,\mathcal H]_n$ denotes $n$ nested commutators, e.g. $[S,\mathcal H]_2=[S,[S,\mathcal H]]$. For $\tilde{\mathcal H}$ to be block diagonal we demand that:
    \begin{equation}
    \label{eq:SW_constraint}
        \mathcal H_1-i[S,\mathcal H_0]=0.
    \end{equation}
Plugging this constraint into the series expansion of $\tilde{\mathcal H}$ yields:
\begin{equation}
\begin{aligned}
    \tilde{\mathcal H} &=\mathcal H_0 +\sum_{n=2}\frac{(n-1)(-i)^{n-1}}{n!}[[S,\mathcal H_1]]_{n-1}\\
    &=\mathcal H_0-\frac i2[S,\mathcal H_1]-\frac13[S,[S,\mathcal H_1]]+\frac{i}{8}[S,[S,[S,\mathcal H_1]]]+\ldots.
    \end{aligned}
\end{equation}
To obtain the exact form of $S$ we introduce the definitions of $\mathcal H_0$, $\mathcal H_1$ into Eq.~(\ref{eq:SW_constraint}) and multiply by the projectors $P_\mu,P_\nu$ on both sides:
\begin{equation}
    \begin{aligned}
        P_\mu \mathcal H_1P_\nu(1-\delta_{\mu\nu})-iP_\mu SP_\nu \mathcal H_0P_\nu+iP_\mu \mathcal H_0 P_\mu SP_\nu=0.
    \end{aligned}
\end{equation}
This gives two equations depending on whether $\mu=\nu$ or $\mu \neq \nu$:
\begin{align}
    \mu&=\nu:\quad P_\mu \mathcal H_0 P_\mu SP_\mu=P_\mu S P_\mu \mathcal H_0 P_\mu \Longrightarrow P_\mu S P_\mu = \gamma P_\mu,\\
    \mu&\neq\nu: \quad iP_\mu \mathcal H_1P_\nu=P_\mu \mathcal H_0 P_\mu SP_\nu-P_\mu SP_\nu \mathcal H_0P_\nu. \label{eq:projection}
\end{align}

The first case yields a trivial solutions, so we will focus on the second one. At low energies only the excitations by a single sector ($\nu=\mu\pm1$) are relevant, so Eq.~(\ref{eq:projection}) takes a form:
\begin{equation}
    iP_\mu\mathcal H_1P_{\mu+1}=P_\mu \mathcal H_0 P_\mu SP_{\mu+1}-P_\mu SP_{\mu+1}\mathcal H_0P_{\mu+1}.
\end{equation}
This equation is most easily solved iteratively, assuming that for sufficiently large $n$ the following equality holds:
\begin{equation}
    P_\mu S^{(n)}P_{\mu+1}=(-iP_{\mu}\mathcal H_1P_{\mu+1}+P_\mu\mathcal H_0P_\mu S^{(n-1)}P_{\mu+1})(P_{\mu+1}\mathcal H_0P_{\mu+1})^{-1}.
\end{equation}
Starting with $P_\mu S^{(0)}P_{\mu+1}=0$ we get:
\begin{equation}
    P_\mu S^{(1)}P_{\mu+1}=-i\frac{P_\mu \mathcal H_1 P_{\mu+1}}{P_{\mu+1}\mathcal H_0P_{\mu+1}}.
\end{equation}
Iterating further we arrive at:
\begin{equation}
\begin{aligned}
    P_\mu S^{(\infty)}P_{\mu+1}&=-i\Big(1+\frac{P_\mu \mathcal H_0P_\mu}{P_{\mu+1}\mathcal H_0P_{\mu+1}}+\frac{(P_\mu \mathcal H_0P_\mu)^2}{(P_{\mu+1}\mathcal H_0P_{\mu+1})^2}+\ldots\Big)\frac{P_\mu \mathcal H_1 P_{\mu+1}}{P_{\mu+1}\mathcal H_0P_{\mu+1}}\\
    &=-i\frac{P_\mu \mathcal H_1 P_{\mu+1}}{P_{\mu+1}\mathcal H_0P_{\mu+1}-P_\mu\mathcal H_0P_\mu}.
    \end{aligned}
\end{equation}
The expressions in the denominator will be treated as numbers, taken to be energy difference between the excited state considered in the specific term and the ground state energy (they are all of order $\Delta$ but depend on the concrete process being considered, e.g. transition of a hole from an $\alpha$ orbital into a singly occupied $d$ orbital raises the energy by $U-\Delta_\alpha$). They will be denoted $\tilde U=P_1\mathcal H_0P_1-P_0\mathcal H_0P_0$ and $\tilde U'=P_2\mathcal H_0P_2-P_1\mathcal H_0P_1$.
This expression for the projections of $S$ can be then plugged into the Baker-Campbell-Hausdorff formula, yielding (up to the fourth order):
\begin{equation}
    P_0 \mathcal H P_0 =P_0 \mathcal H_0 P_0- \frac{1}{\bar U}P_0\mathcal H_1P_1\mathcal H_1P_0 + \frac{1}{\tilde {U'}^3}(P_0\mathcal H_1P_1\mathcal H_1P_0\mathcal H_1P_1\mathcal H_1P_0-\frac12P_0\mathcal H_1P_1\mathcal H_1P_2\mathcal H_1P_1\mathcal H_1P_0).
\end{equation}
This is the form of the Schrieffer-Wolff transformation used throughout this letter; it can be found in several works, {\it cf.}~Ref.~\cite{Kadzielawa-Major_2014}. As stated in~\cite{Bravyi_2011} the series expansion of the Hamiltonian presented above converges absolutely if $\Delta/16>||V||$ (assuming the low energy sector has negligible width). Note, however, that the derivations presented in this letter do not obey the latter inequality, having the ratio of $\Delta/||V||\sim5-10$, and thus the results should be treated carefully.
\FloatBarrier

\section{Derivation of the $d$ band exchange interaction}
The effective interaction between the holes located in the $d$ orbitals can be obtained by expanding the Emery Hamiltonian up to the fourth order in $U$. The Hamiltonian resulting from this expansion reads:
\begin{equation}
\label{eq:H_eff_epsilon}
    \begin{aligned}
    H_{4^{\mathrm{th}}}&=
    \frac{1}{(U-\Delta_\alpha)^3}\Bigg\{ \sum_{\substack{j_1\neq j_2,j_3\\j_4,j_5,\sigma}}T_{j_3j_1}T_{j_1j_4}T_{j_4j_2}T_{j_2j_5}\left[\alpha_{j_3\sigma}^\dag (\tilde n_{j_1\Bar{\sigma}}\tilde n_{j_2\Bar{\sigma}}
    + S^{\Bar{\sigma}}_{j_1}S^{\sigma}_{j_2})\alpha_{j_5\sigma}\right.\\
    &\left.-\alpha_{j_3\sigma}^\dag(\tilde n_{j_1\Bar{\sigma}}S^{\Bar\sigma}_{j_2}+\tilde n_{j_2\sigma}S^{\Bar{\sigma}}_{j_1})\alpha_{j_5\Bar\sigma}\right]\\
    &+\sum_{\substack{j_1,j_2,j_3\\j_4,j_5,\sigma}}T_{j_2j_1}T_{j_1j_3}T_{j_4j_1}T_{j_1j_5}(\alpha_{j_2\sigma}^\dag \alpha_{j_3\Bar\sigma}\alpha_{j_4\Bar\sigma}^\dag \alpha_{j_5\sigma}+\alpha_{j_2\sigma}^\dag \alpha_{j_3\sigma}\alpha_{j_4\sigma}^\dag \alpha_{j_5\sigma})\tilde n_{j_1\Bar\sigma} \Bigg\},
    \end{aligned}
\end{equation}
where $T_{ij}=2t_{pd}\mu({\bf r}_i-{\bf r}_j)$ and $\Delta_\alpha=\Delta-2t_{pp}\nu(0)$ are the shorthand notation for the parameters of the Hamiltonian in the $\{ d, \alpha, \beta \}$ basis.
This Hamiltonian has many different terms, but most of them disappear while performing further projection into lower and lower energy subspaces because of the requirement that both $d$ and $\alpha$ bands are half-filled. This excludes all processes that move holes between the $\alpha$ orbitals, only allowing for exchange interactions. The remaining terms are:
\begin{equation}
\begin{aligned}
H_{dd}&=\frac{1}{(U-\Delta_\alpha)^3} \Bigg\{ \sum_{\substack{j_1\neq j_2,j_3\\j_4,\sigma}}T_{j_3j_1}T_{j_1j_4}T_{j_4j_2}T_{j_2j_3} (\tilde n_{j_1\Bar{\sigma}}\tilde n_{j_2\Bar{\sigma}}
    +S^{\Bar{\sigma}}_{j_1}S^{\sigma}_{j_2})n^\alpha_{j_3\sigma}-(\tilde n_{j_1\Bar{\sigma}}S^{\Bar\sigma}_{j_2}+\tilde n_{j_2\sigma}S^{\Bar{\sigma}}_{j_1})s^{\alpha\sigma}_{j_3}\\
    &+\sum_{\substack{j_1,j_2,j_3\\j_4,j_5,\sigma}}T_{j_2j_1}T_{j_1j_3}T_{j_3j_1}T_{j_1j_2}(n^\alpha_{j_2\sigma}(1-n^\alpha_{j_3\Bar\sigma})+n^\alpha_{j_2\sigma})\tilde n_{j_1\Bar\sigma} \Bigg\}.
    \end{aligned}
\end{equation}
Of all the terms the first one is the most important, as it couples different $d$ sites with each other as the only one in the Hamiltonian (the rest couples the $d$ and $\alpha$ orbitals with amplitudes being much smaller than the second order processes discussed in the main text). This exchange interaction can further be rewritten to show that it does not depend on the spin of the hole on the $\alpha$ site, and it takes the form:
\begin{equation}
    H_{dd}=\frac{1}{(U-\Delta_\alpha)^3}\sum_{(j_1,j_2),j_3,j_4}2T_{j_3j_1}T_{j_1j_4}T_{j_4j_2}T_{j_2j_3}({\bf S}_{j_1}{\bf S}_{j_2}+\frac{n_{j_1}n_{j_2}}{4})\approx\frac{1}{(U-\Delta_\alpha)^3}\sum_{(j_1,j_2)}4T_{j_2j_1}^2T_{j_1j_1}^2({\bf S}_{j_1}{\bf S}_{j_2}+\frac{n_{j_1}n_{j_2}}{4}),
\end{equation}
where the sum over $(j_1,j_2)$ denotes summation over pairs of indices $j_1$, $j_2$ (this should be distinguished from the summation over the nearest neighbour pairs).
In order to obtain the value of $J_{dd}$, summation should be performed over all possible values of the ligand $p$ orbital indices $j_3$ and $j_4$ -- see Eq.~\eqref{eq:Jdd} below for the final result. Note that this is a daunting task which, however can be accurately approximated by only including the processes where at least one ligand is at the same site as the $d$ orbital. This is because the on-site $d-\alpha$ hopping is an order of magnitude stronger than farther range hopping. 

Another contribution to the exchange interaction can be obtained by continuing the expansion in $U_{pp}$ up to the fourth order. The spin exchange term is identical (up to the constant terms) with a prefactor proportional to $\propto \big[\psi_{0000}U_{pp}+\Delta-2t_{pp}\nu(0)\big]^{-3}$.
Even those two contributions together still result in a relatively small magnitude of the $J_{dd}$ exchange interactions, as showcased in Table~II of the main text.

\section{Parameters of the effective Hamiltonian}
\label{app:coefficients}
The parameters of the effective ancilla Hamiltonian can easily be related to the parameters of the original Emery model and the functions $\mu$, $\nu$, $\chi$ and $\psi$ with the following relations, stemming from the consecutive Schrieffer-Wolff transformations:
\begin{align}
J^{\alpha\alpha}_{j_1j_2}&=\frac{8(t_{pp}\nu(\textbf{r}_{j_1}-\textbf{r}_{j_2}))^2}{\psi_{0000}U_{pp}}=J^{\beta\beta}_{j_1j_2},\\
    J^{d\alpha}_{j_1j_2}&=8(t_{pd}\mu(\textbf{r}_{j_1}-\textbf{r}_{j_2}))^2\bigg(\frac{1}{\psi_{0000}U_{pp}+\Delta-2t_{pp}\nu(0)}+\frac{1}{U-\Delta+2t_{pp}\nu(0)}\bigg),\\
    J^{\alpha\beta}_{j_1j_2}&=8(t_{pp}\chi(\textbf{r}_{j_1}-\textbf{r}_{j_2}))^2\bigg(\frac{1}{\psi_{0000}U_{pp}+4t_{pp}\nu(0)}+\frac{1}{\psi_{0000}U_{pp}-4t_{pp}\nu(0)}\bigg),\\
    \tau^{\alpha}_{j_1j_2j_3}&=\frac{4t_{pp}^2\chi(\textbf{r}_{j_1}-\textbf{r}_{j_2})\chi(\textbf{r}_{j_2}-\textbf{r}_{j_3})}{4t_{pp}\nu(0)},\\
    \bar\tau^\alpha_{j_1j_2j_3}&=\frac{4t_{pp}^2\chi(\textbf{r}_{j_1}-\textbf{r}_{j_2})\chi(\textbf{r}_{j_2}-\textbf{r}_{j_3})}{\psi_{0000}U_{pp}-4t_{pp}\nu(0)},\\
    \tau^\beta_{j_1j_2j_3}&=\frac{4t_{pp}^2\nu(\textbf{r}_{j_1}-\textbf{r}_{j_2})\nu(\textbf{r}_{j_2}-\textbf{r}_{j_3})}{\psi_{0000}U_{pp}},\\ \label{eq:Jdd}
    J^{dd}_{j_1j_2j_3}&=32t_{pd}^4\mu(\textbf{r}_{j_1}-\textbf{r}_{j_2})^2\mu(\textbf{r}_{j_2}-\textbf{r}_{j_3})^2
    \Bigg(
    \frac{1}{\big(U-\Delta+2t_{pp}\nu(0)\big)^3}+\frac{1}{\big(\psi_{0000}U_{pp}+\Delta-2t_{pp}\nu(0)\big)^3}\Bigg).
\end{align}
\end{document}